\documentclass[11pt]{article}
\usepackage{graphicx,makeidx}
\usepackage{amsmath}                                  
\usepackage{epsfig}
\usepackage{amssymb} 
\usepackage{color}

\newcommand{\bea}{\begin{eqnarray}}
\newcommand{\eea}{\end{eqnarray}}

\makeindex 
\begin{document}
 
\title{Optimization by  Record Dynamics.  
}
\author{Daniele Barettin\\
Dept. of Electronics Engineering, University of Rome, Tor Vergata, Rome (Italy)\\
 and \\Paolo Sibani  \\
 FKF, University of Southern Denmark, Odense (Denmark)
 } 
\date{}
\maketitle
\abstract{

\noindent Large  dynamical changes  in
thermalizing   glassy systems 
are triggered  by  trajectories  crossing  record sized barriers,
a behavior  revealing the presence of  a hierarchical structure in
configuration space. The observation  
 is here turned into  a  novel  local
search optimization algorithm 
dubbed  Record Dynamics Optimization, or RDO.
RDO uses the Metropolis  rule  to  accept or reject
candidate solutions depending on the value of a parameter
akin to the  temperature, and 
 minimizes the  cost function of the problem at hand 
  through   cycles where 
its  `temperature' is raised and subsequently decreased
in order to expediently generate record high (and low) values
of the cost function.
Below, RDO is introduced  and then  tested 
 by   searching   the ground state of the Edwards-Anderson spin-glass model, 
in two and three spatial dimensions. 
A popular and  highly efficient  optimization algorithm,
Parallel Tempering (PT) is applied to the same problem as a benchmark.
 RDO and PT turn out to produce solution of similar quality for similar numerical effort, but 
 RDO is simpler to program and additionally yields   geometrical information on the
system's  configuration space which is  of interest in many applications.
In particular, the effectiveness of  RDO  strongly  indicates  the presence of the above mentioned
hierarchically
organized configuration space, with metastable regions   indexed by the cost (or energy) 
the transition states connecting  them.
}

\section{Introduction} 
 Built on analogies with physical or biological processes, heuristic optimization techniques
 are widely used in science\cite{Kirkpatrick83,Cerny85,Huang86,Goldberg89,Dueck90,Nourani99,boettcher01}. 
 Of present interest   is 
  Simulated Annealing (SA), a well known  local search algorithm based on the Metropolis 
  algorithm, which 
minimizes the cost of candidate solutions in a way
  similar to a physical system minimizing its free energy under cooling~\cite{vanLaarhoven87,Salamon02}. 
  In SA, a proposed  solution  is first generated by locally modifying 
the current solution. Changes lowering the cost are   accepted 
and others are 
 accepted with
  probability $\exp(-\Delta E/T)$, 
  where   $\Delta E  >0$ is the additional cost incurred and where the parameter $T$ 
  is conventionally called temperature. 
  Ideally, a cooling schedule 
gradually decreasing  the temperature  down 
 to zero  should   reach  the ground state, i.e.
   the desired solution of the optimization problem.
  However, in  applications  to  hard combinatorial problems
 SA  invariably gets stuck 
in one of the   many  suboptimal or metastable configurations
which characterize  these systems. Since available local configurational changes 
 mainly get rejected, a larger 
 partial randomization is     required  to obtain further improvements.
 
Large changes leading a thermalizing complex system  from 
a  meta\-stable configuration to another 
are often   triggered by thermal energy fluctuations  of 
record magnitude\cite{Sibani03,Anderson04,Sibani06a,Sibani08}. It is then    
natural to hypothesize that visiting configurations of record-high cost, or energy,  
similarly helps a `thermal' optimization algorithm of the SA type to escape  suboptimal solutions.

The configuration space, or energy landscape, of the Edward-Anderson spin glass~\cite{Edwards75}  was previously investigated  
using Extremal Optimization~\cite{boettcher01}, an optimization and exploration algorithm indifferent to energy  
barriers, and by the Waiting Time Method~\cite{Dall01},
a kinetic Monte Carlo. algorithm with no rejections. 
The analysis led to the conclusion that, in order to achieve a lower  BSF value, which is desirable in 
optimization, the barrier $B(t)$ must  previously  reach  a new high record. Importantly, this property is not 
associated to the algorithms used, but pertains to all energy landscapes which can be 
 coarse-grained into inverted 
binary trees where  nodes represent metastable configurations~\cite{Hoffmann88}
and height represents the energy.
 Motivated by the above considerations, 
 the Record Dynamics Optimization (RDO) algorithm  introduced  below   
   dynamically generates  a non-monotonic SA  schedule 
where    heating and cooling phases alternate. 
Each heating phase terminates once a record high  `barrier'
(defined below) is encountered  and each cooling phase terminates once a 
state of record low cost is found.
M\"{o}bius et al.\cite{Mobius97} earliere introduced a 
non monotonic  annealing schedule where 
temperature
oscillations are  controlled by a tunable  parameter
instead of being determined by 
intrinsic  geometrical properties of the landscape.
 
 For demonstration purposes, RDO  is used 
 to  search for the ground state of a three dimensional Edwards-Anderson (EA) spin glass~\cite{Edwards75},
 a standard NP hard optimization problem. For completeness,
 it is further  applied to the two dimensional EA model. 
RDO   performance   is then compared to  that of a carefully 
 optimized version of Paralle Tempering (PT). 
The numerical effort  needed 
to obtain results  of comparable  quality is similar for the two methods.
However, RDO   has fewer   tunable
parameters and is more easily  implemented.
 Secondly, RDO  provides, at no extra cost,   
some information on  configuration space structure
which might be of interest in  landscape explorations.

\section{The RDO algorithm} First some notation: 
A sweep in a MC run comprises  a number of elementary moves or queries, i.e. the 
generation and acceptance or rejection of a candidate move, equal to
the number of independent variables of the problem. The number of sweeps 
carried out up to a certain point 
is dubbed time and denoted by the symbol $t$. 
Each query generates a putative solution or state, and 
the ordered sequence of states sampled   in  $[0, t]$ 
is called a trajectory. The cost
associated to a state is called its energy  $E$. 
The Best So Far energy, ${\rm BSF}(t)$,
 is the lowest energy sampled in a single trajectory in  
$[0,t]$. The barrier $B(t)$ associated to a state 
sampled at time $t $ is $B(t) \stackrel{\rm def}{=} E(t)-{\rm BSF}(t)$. Lower case symbols are used for quantities scaled by
system size, i.e.  in the example considered 
 $b(t)$ is the barrier energy per spin.
We stress that  the BSF and barrier functions   are stochastic processes 
 and that inherent geometrical  properties of the landscape
 can only be estimated  by  averaging over a suitably large 
 ensemble of  independent trajectories.

The RDO algorithm  comprises an initial phase  followed by  a succession of  cooling and a heating phases
controlled by record events. 
  Each  of these  phases involves 
decreasing  or increasing the temperature  within a set of 
of    $22$ predefined and equidistant values in the temperature  range   $[T_{MIN}, T_{MAX}]$.
Several preliminary simulations showed that  
   $T_{MAX}= 1.2$, slightly above the critical temperature of the 3d model is a good choice. 
Furthermore, $T_{MIN}= 0.3$  was chosen as  BSF values are   rarely, if at all,  found  
below $T=0.3$, 
We let the system cool and heat \emph{ad libitum} since each cooling or heating  phase 
produces gradually lower  extremal  values. Once the minimal temperature is reached, and no further BSF is found,
the algorithm stops. 
\begin{enumerate}
\item Initialization of BSF and barrier  values: Any short ÔnaiveÕ \cite{Salamon02} optimization at a constant 
temperature typically slightly below the critical temperature $T_g$ will produce the first BSF value, 
${\rm BSF}_0$. The first high barrier value $b_0 >{\rm BSF}_0$  is found by running the algorithm at a slightly 
higher constant temperature. For $i = 1,2 \ldots$, 
 the `barrier'  $ B(t) = E(t) -BSF_i $ is used to control the algorithm.
 The highest barrier overcome in heating phase $i$ is called  $B_i$.
\item Cooling: Let  $S_{B,i}$	be the  configuration  corresponding to  
$B_i$.
 Starting from $S_{B,i}$ run SA with decreasing temperature until a lower BSF 
value is found.
If no lower BSF is found, cooling stops after 
     $N_{\rm step}=50000$ sweeps. 
\item  Running at constant T: the Metropolis algorithm
at constant T is used  until either $m$  new BSF values have
been found or the  preset max time is exceeded. In practice $m$  is a small integer,
i.e. $m=3$ in the present simulations.
 This step ensures  that once the correct region of configuration
space is identified,  some time is spent 
exploring it. ${\rm BSF}_{i+1}$  is the lowest BSF value identified 
during  this	phase.
\item Heating: starting from $S_{i+1}$, the configuration corresponding to
  BSF$_{i+1}$,
 heat the system until  $B(t) = E(t) - {\rm BSF}_{i+1} > B_{i}$.
The achieved record value of $B(t)$ defines $B_{i+1}$.
\item Set $i+1\rightarrow i$, go to step 2 and repeat \emph{ad libitum}.
\end{enumerate}
\section{Parallel tempering}
Parallel Tempering (PT)  avoids trapping   by
independently searching  a number $N_T$of identical replicas of the problem
at hand.  The $m$'th replica is explored by a conventional Metropolis 
algorithm run at  a  temperature $T_m$. 
Additional  configurational   swaps between replicas,
also  controlled by the Metropolis  criterion in order
 to ensure detailed balance,
provide the sought  escape route from suboptimality.
A successfull PT implementation requires consideration of the temperatures at which 
the replicas are run and a compromise between the number of attempted swaps 
and the number of standard queries within the replicas.
The reader is referred to~\cite{Marinari,Hukushima,Moreno} for a in-depth
discussion of PT.  The brief summary provided below describes the 
implementation presently used to benchmark  RDO. 

\begin{enumerate}
\item  $N_T$ different copies of the system
are  updated  in parallel  at  temperatures
 $T_{m} > T_{m+1}, \quad m = 1...N_T$  through one or more Monte Carlo
 sweeps.
\item A proposed  swap between
   configuration $ C_m $ and  
   $C_ {m +1} $ is accepted or rejected according to the Metropolis criterion.
   Defining  $\beta_m = 1 / T_m$, and 
 \bea
\Delta S = \biggl[ \beta_{m + 1}E(C_m) + \beta_m
  E(C_{m +1}) \biggr] -  \biggl[ \beta_m E(C_m) +
  \beta_{m + 1}E(C_{m +1}) \biggr],
\eea  
   the exchange is accepted with 
 probability  $ \min(1,e^{- \Delta S}) $.
\item Further exchanges between  the configurations  associated with 
$\beta_{m + 1}$ and  $ \beta_{m +2} $ are accepted or rejected in the same
way, eventually exploring the whole set of temperatures.
\item Go to step 1 and repeat \emph{ad libitum}.
\end{enumerate}

After a number of exploratory  simulations,  the highest  temperature was chosen as  $T_{max} = 1.6$, 
a value higher than the critical temperature of the Edwards-Anderson spin glass i.e.
$T_c\approx 0.95$ \cite{Marinari2}. The lowest temperature is dynamically determined as discussed 
below.
A suitable number of temperatures for PT  is generally estimated to be  $N_T \approx \sqrt{N_{\rm spin}}$\cite{Hukushima}. 
In the following,  $N_T=30,50$ and $90$ are used for $L=30,50$ and $100$  in the 2d simulations, while 
$N_T=30,40$ and $80$ are used in  the 3d  case  for $L=8,14$ and $20$.

To  accept an  exchange between 
 copies with probability   $\approx 0.5$,  a value considered to be optimal~\cite{Moreno}, 
  the $T_{m}$ values are  treated as  dynamical 
variables using the  recursive method 
described in Ref.\cite{Hukushima}. Initially, the inverse temperatures $\beta_m$ are set to
\bea
\beta_m=\beta_1+ (\beta_M-\beta_1) \frac{m-1}{M-1}
\eea
with $M=N_T$. 
The  updated   set $ \{\ \beta_m' \}\ $ is obtained using  the sampled exchange rates
$p_m$ between configurations at inverse temperatures $\beta_m$ and $\beta_{m-1}$:
$$\beta_1'= \beta_1$$
$$\beta_m'=\beta_{m-1}'+(\beta_m - \beta_{m-1})\ \frac{p_m}{c} \quad \mbox{with } m=2,...,M$$. 
\bea
c=\frac{1}{M-1}\sum_{m=2}^M p_m \label{dTnum}
\eea
While in Ref.\cite{Hukushima}   temperatures are only updated initially 
to reach  the constant  values used in the  simulation, we found it  more 
convenient to update  them during the simulation itself, at
at logarithmically equidistant
times  $2^n \times 100$ MC sweeps, with $n=1,2,..N$.

Two different benchmarks for  RDO  are provided. The first, our `fast' PT, has  $N=10$  
and   $N_{\rm step}=102400$ sweeps per replica. Adding the computational effort
for all replicas, PT  is
eight time faster than  RDO but   produces results of somewhat lesser  quality.
The second version, `slow'  PT,  has  $N=13$  
and $N_{\rm step}=819200$,  
with the  total number of sweeps  approximately corresponding to that used in 
our  RDO implementation. 
Both versions of the  PT algorithm  include a final quench to $T=0$, a step omitted  in  
RDO. 
Importantly, the  PT versions implemented are carefully  optimized and  
based on   the recent  literature on the subject.  

\section{Results}
\subsection*{The model}
The  EA model\cite{Edwards75} with Gaussian interactions deals with a set of  spins $\sigma_i=\pm 1$
which are  placed on a d-dimensional discrete grid with  linear size $L$ and 
periodic boundary conditions. 
The spins interact via a coupling matrix   $\mathbf J$ which 
 is symmetric,  has diagonal elements  all equal  zero
and   off-diagonal elements  $J_{ij}$ likewise  equal zero,  unless  spins
$i$ and $j$ reside  on neighboring grid points. In this
case,  and for $i<j$,   their values are  independently drawn from a Gaussian distribution with  
zero average and unit variance. The energy of a  configuration $\alpha$ is then
given by 
\begin{equation}
E^\alpha = \frac{1}{2} \sum_{i,j} J_{ij} \sigma_i^\alpha  \sigma_j^\alpha.
\end{equation}
Determining the energy of the ground state of the 3d Edwards-Anderson spin
glass problem in its different guises  is an NP hard combinatorial
problem which has  been attacked 
using a variety of techniques. For future reference we note that  Pal~\cite{Pal96}  combined 
 a genetic alghorithm with local search and found that the ground state energy per spin is
$e_{gs} = -1.699926 +2.1373 L^{-3}$, where the first term is the thermodynamic limit and
the second term describes finite size corrections. More recently, Rom\'{a} et al.~\cite{Roma09} used 
Parallel Tempering and considered three different finite size corrections
to the thermodynamic limit, one of which is  $e_{gs} = -1.7000 + 2.01 L^{-2.94}$.
The same authors find $e_{gs} = -1.3149 + 1.3 L^{-2.28}$ for the 2d system.

In this work, the EA  model is used to test  how  RDO works and
to benchmark it  against PT. As there is no ambition  to improve on
 existing estimates of the model's ground state energy, we 
 limit ourselves to three different system sizes in both 2d and in 3d, and offer
no analysis of  finite size corrections. 
Most  results  are  averages
over $N_{\rm sample}$ of independent realizations of  $\mathbf J$ and, 
unless otherwise  specified, the combination $N_{\rm sample}=200, 50$ and $10$ is
either used  for  2d systems
of linear size $L=30, 50$ and $100$ or  for 3d systems of linear size 
$L=8, 14$  and $20$.
Since the RDO algorithm produces trajectories with varying number of steps and 
with BSF$(t)$ values obtained at different times,  coarse-graining each trajectory  is
required  to average different trajectories.
Accordingly, within each trajectory, the  measured values 
 were averaged  every 750000 steps, as long as possible.
 This choice fits the worst case encountered for every size studied, i.e. provides   
a partition  of the trajectory with the slowest decay of the BSF energy.
Faster trajectories were padded  with the last BSF energy value achieved.
 In  this way, the computational time used by  the RDO algorithm to achieve a certain averaged
BSF energy value is slightly overestimated.

\subsection*{Temperature,  barriers and energies in  RDO}
Since alternating heating and cooling phases are characteristic features
of the algorithm, the time dependence of the
temperature  and the related time dependence of the record barriers 
which control the switch from heating to cooling are discussed first.
Second, we discuss the time dependence of the BSF energy per spin, which,
at each stage of the calculation, provides the RDO estimate of the ground state energy.
Finally, the randomizing effect of a temperature cycle is discussed   in terms of  Hamming
distances. 

\begin{figure}
 $
\begin{array}{cc}
\rotatebox{-90}{
\includegraphics[width=0.33\linewidth]{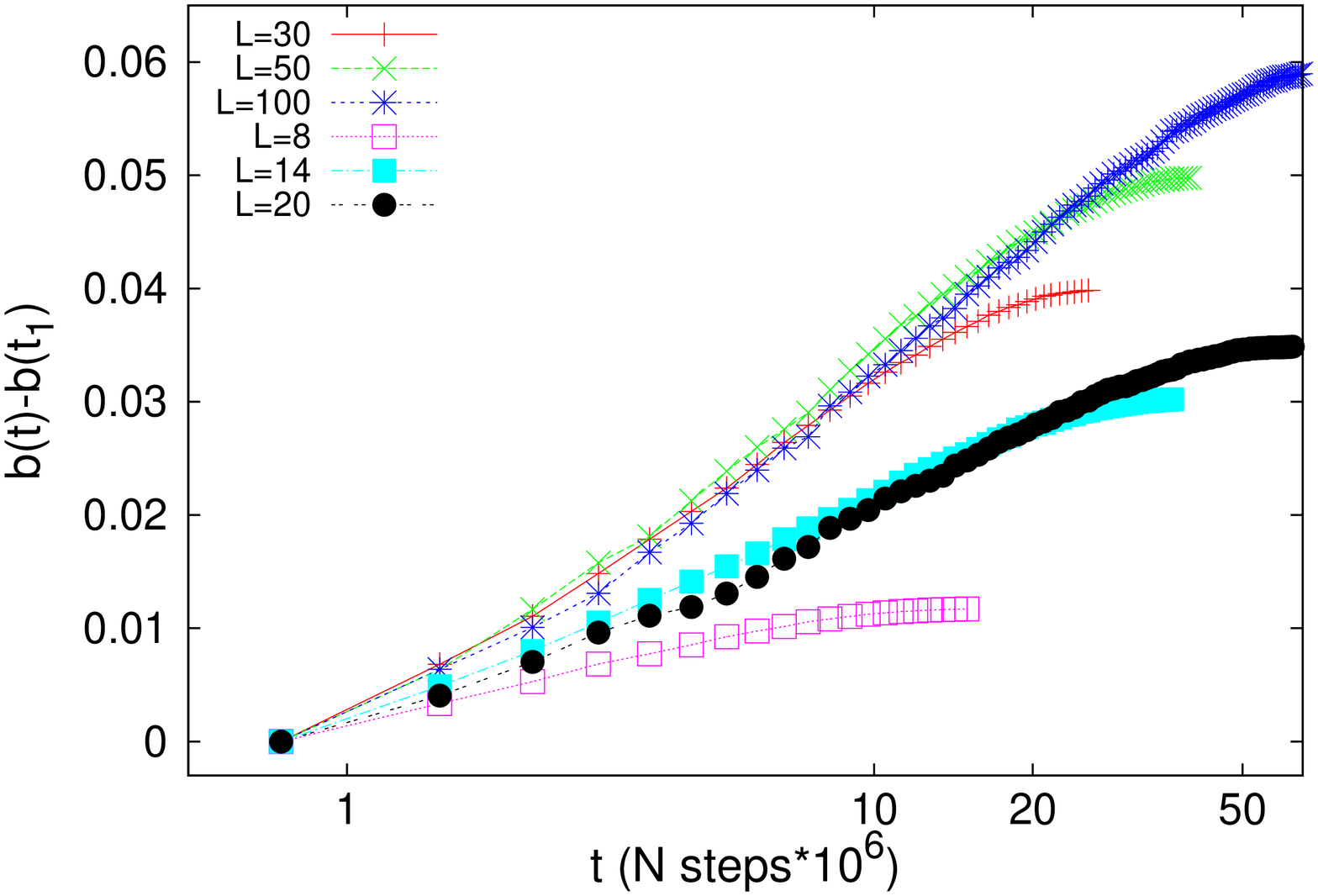} } & 
\rotatebox{-90}{\includegraphics[width=0.33\linewidth]{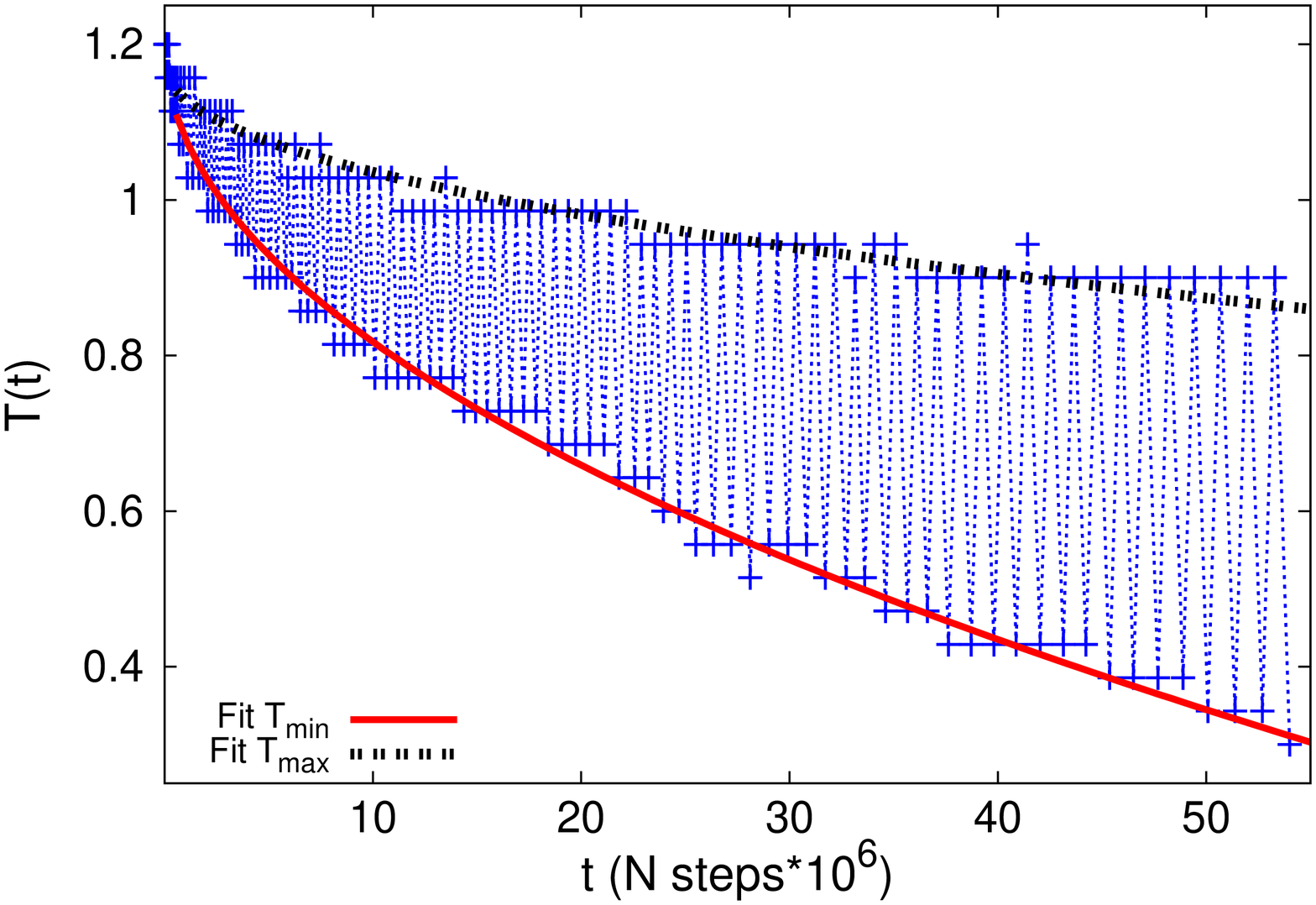}}
\end{array}
$ 
\caption{(Color online) 
Left:
 the average barrier per spin $b(t)$ with  the initial value 
subtracted and for all the systems considered, is  plotted versus time on a   logarithmic
 horizontal scale. 
Right: the  temperature $T(t)$ vs. time for 
a single trajectory of  a 2d system of linear size $L=100$. The lower and upper  
lines show  fits to  the local minima and maxima $T_{\rm min}(t)$ and 
$T_{\rm max}(t)$, respectively. 
}
\label{B_T_step} 
\end{figure}

 The left panel  of Fig.~\ref {B_T_step} depicts  the average barrier value per spin, $b(t)$, with the 
initial value subtracted. The trend is in all cases logarithmic, but  the  slope 
is lesser   in 3d than in 2d. The data collapse obtained for sufficiently large sizes shows that
barriers are extensive,  in contrast to barriers   reached  in an \emph{isothermal} relaxation process,
which are sub-extensive\cite{Dall03}. Hence,  heating the system up  is a far more efficient
way  to  partial configurational randomization   than  isothermal relaxation.

\begin{figure}
 $
\begin{array}{cc}
\rotatebox{-90}{
\includegraphics[width=0.33\linewidth]{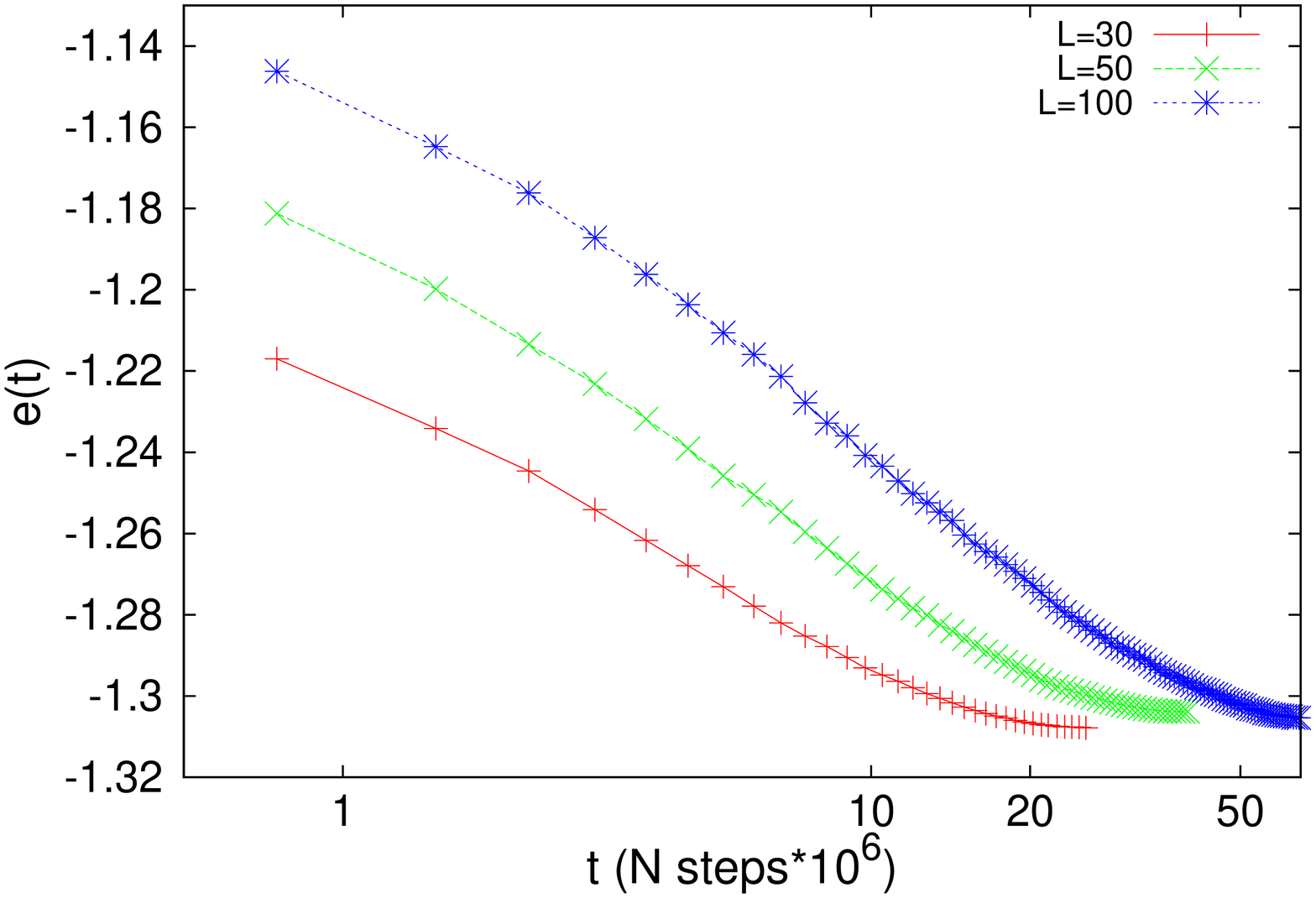} } & 
\rotatebox{-90}{\includegraphics[width=0.33\linewidth]{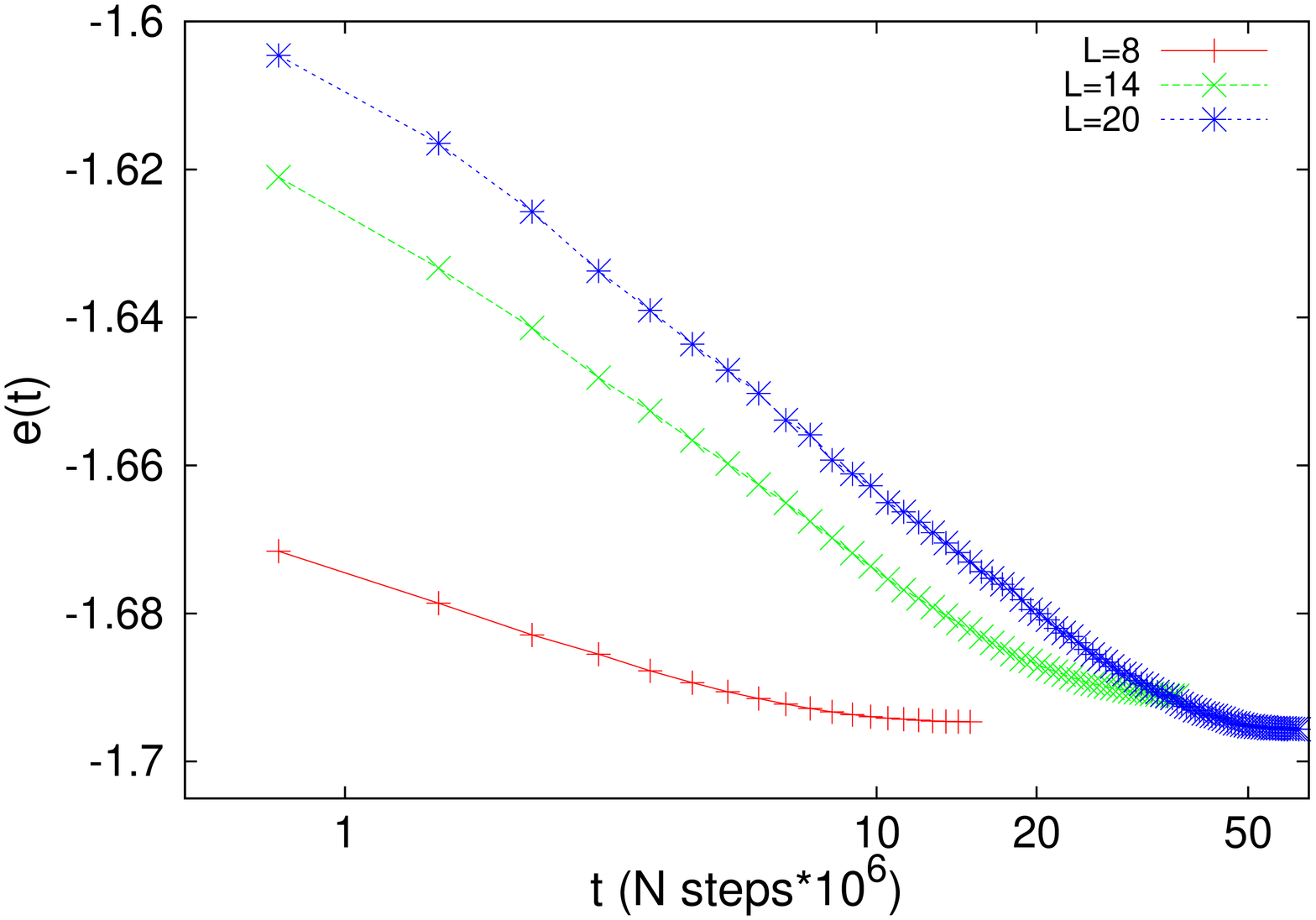}}
\end{array}
$ 
\caption{(Color online) The disorder averaged BSF energy per spin $e$ is
plotted vs.  time for 2d (left panel) and 3d  systems
 using a logarithmic horizontal scale. 
}
\label{Energia_step} 
\end{figure}

In  the right panel of  Fig.~\ref{B_T_step},  the algorithm's  temperature $T$ is plotted vs.
time for  a single  2d system  with $L=100$. Similar 
behavior is observed  for all other systems considered. 
In the curve, each local  minimum   corresponds to the  temperature $T_{\rm min}(t)$ 
for which  a new BSF energy value is found at stage  $3$  of the RDO algorithm,
while each local  maximum corresponds
to a temperature $T_{\rm max}(t)$ reached  at stage  $4$, for which  a  new record sized  barrier is 
found. The upper and lower envelopes of the curves are 
least square fits of the form  $T_{\rm min}(t)=T(t_0)-a_{\rm min}\sqrt{t}$
and  $T_{\rm max}(t)=T(t_0)-a_{\rm max}\log{(t)}$, 
where $a_{min}$ and  $a_{max}$ are numerical constants.
We first note that even though  the barrier values increase in time,
the (un-shifted) energy of the corresponding barrier states
decreases, i.e. the RDO algorithm explores regions of configuration space 
of  gradually  lower energy.
The decreasing trend of 
$T_{\rm max}(t)$ matches the logarithmic decrease of the energy of the different barrier states,
see Fig.~\ref{Energia_step}.  The square root term in 
$T_{\rm min}(t)$ indicates that the BSF  energy minima belonging to \emph{different} regions of the 
landscape  decrease in a roughly linear fashion from
one region to the next  and are reached via  the  diffusion-like process associated to the 
constant temperature search in the third phase of the RDO algorithm.

\begin{figure}
 $
\begin{array}{cc}
\rotatebox{-90}{
\includegraphics[width=0.33\linewidth]{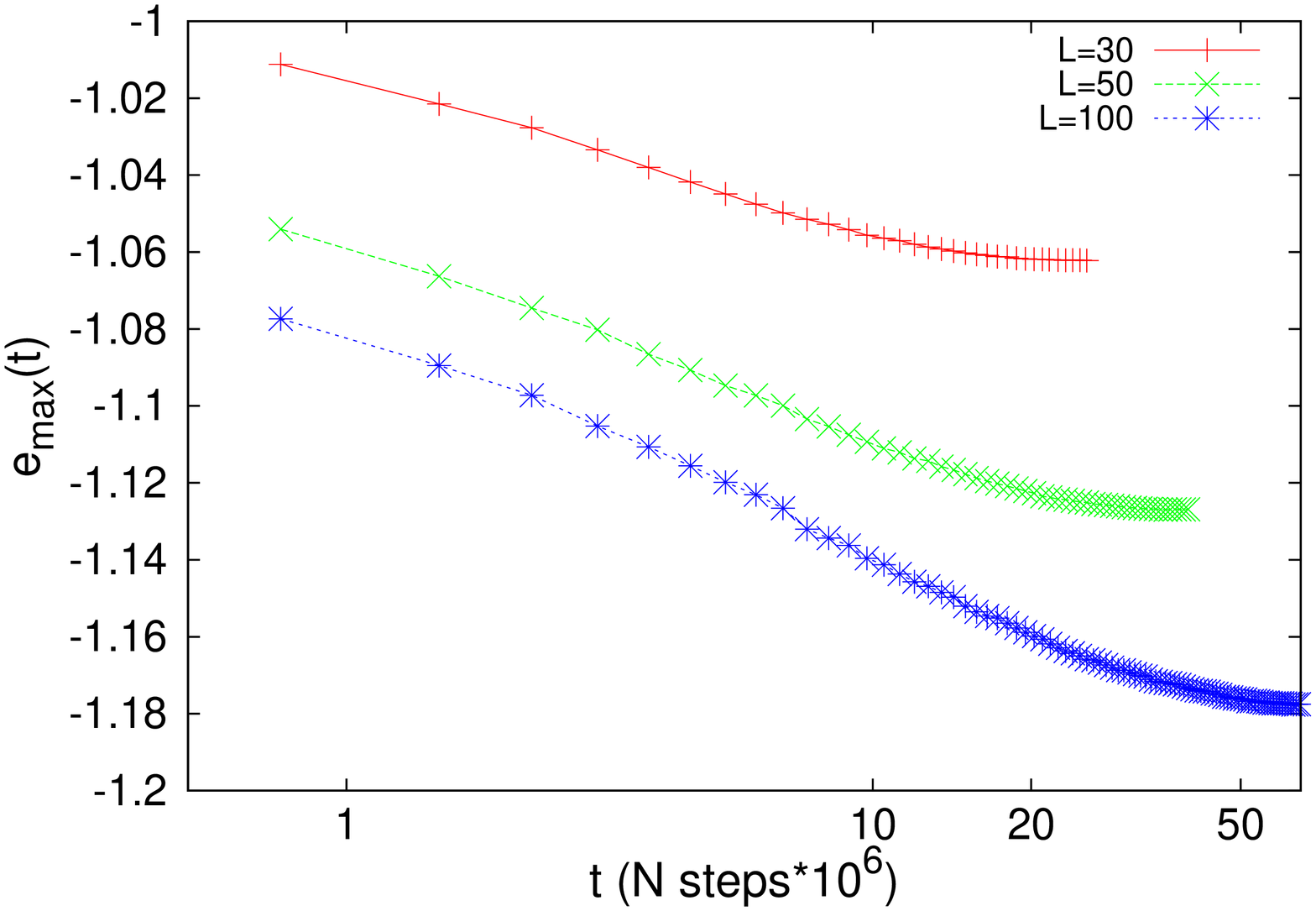} } & 
\rotatebox{-90}{\includegraphics[width=0.33\linewidth]{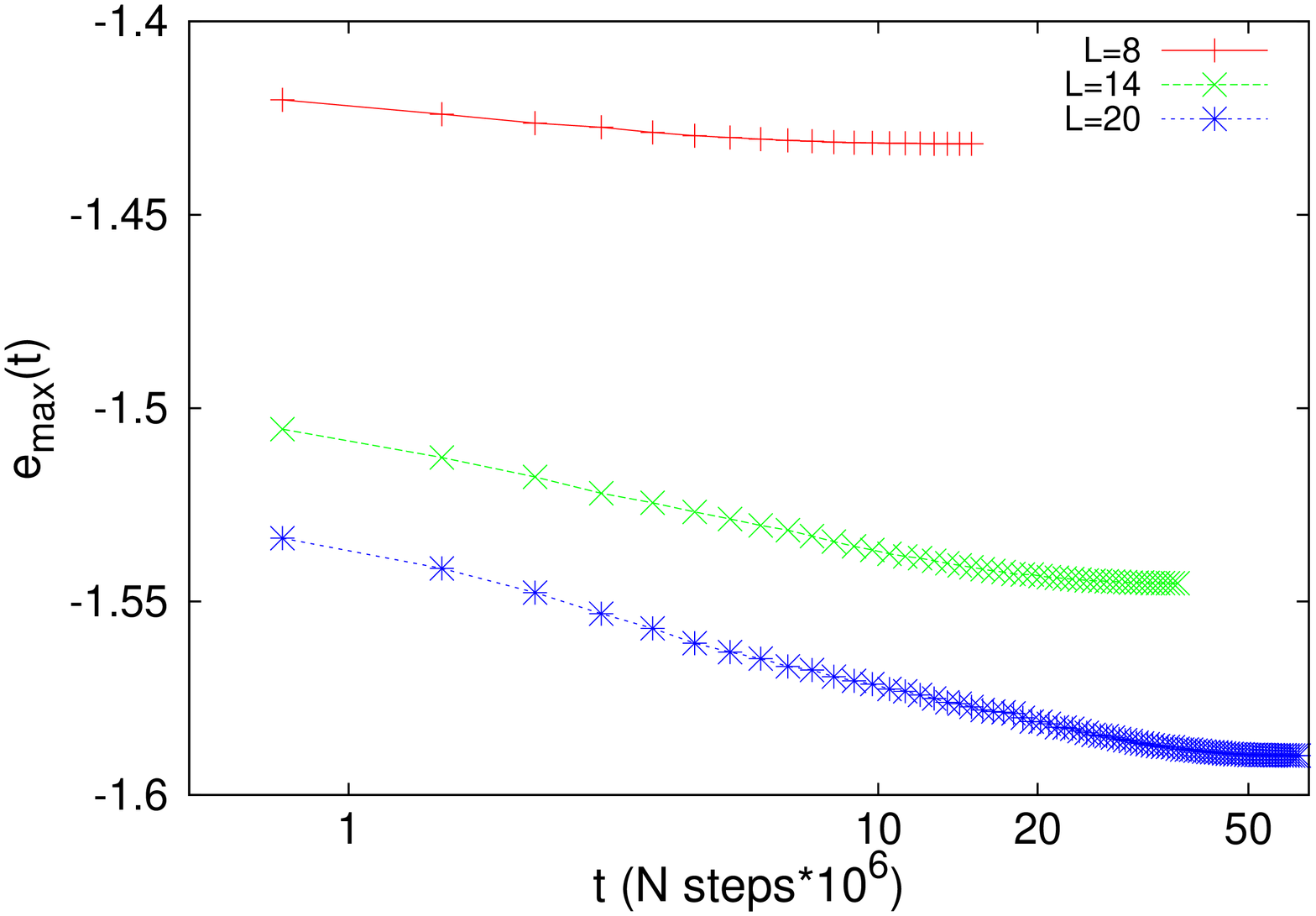}}
\end{array}
$ 
\caption{(Color online) The disorder averaged local energy maxima  $e_{\rm max}(t)$ encountered
 in trajectories are plotted vs. time for the 2d (left) and 3d  systems
using  a logarithmic horizontal scale. 
}
\label{Energiamax_step} 
\end{figure}

All curves show a logarithmic dependence on time.
 The  common asymptotic limit of the curves, which corresponds to the 
 predicted ground state energy value 
   is, as expected,  nearly independent of system size 
and  in agreement with current  numerical estimates~\cite{Roma09}.
In contrast,  the plot of the local energy maxima along 
trajectories which is depicted  Fig.~\ref{Energiamax_step}
shows that  $e_{\rm max}(t) =e(t) + b(t)$
retains a  system size dependence
throughout the simulation.
A glance to Fig.\ref{B_T_step} shows that the latter mainly stems from  the
system size dependence of the initial barrier value.
We stress  that both $e_{\rm max}$ and $e(t)$
decrease logarithmically in  time while  the barrier $b(t)$ 
correspondingly increases.

\begin{figure}
 $
\begin{array}{cc}
\rotatebox{-90}{
\includegraphics[width=0.33\linewidth]{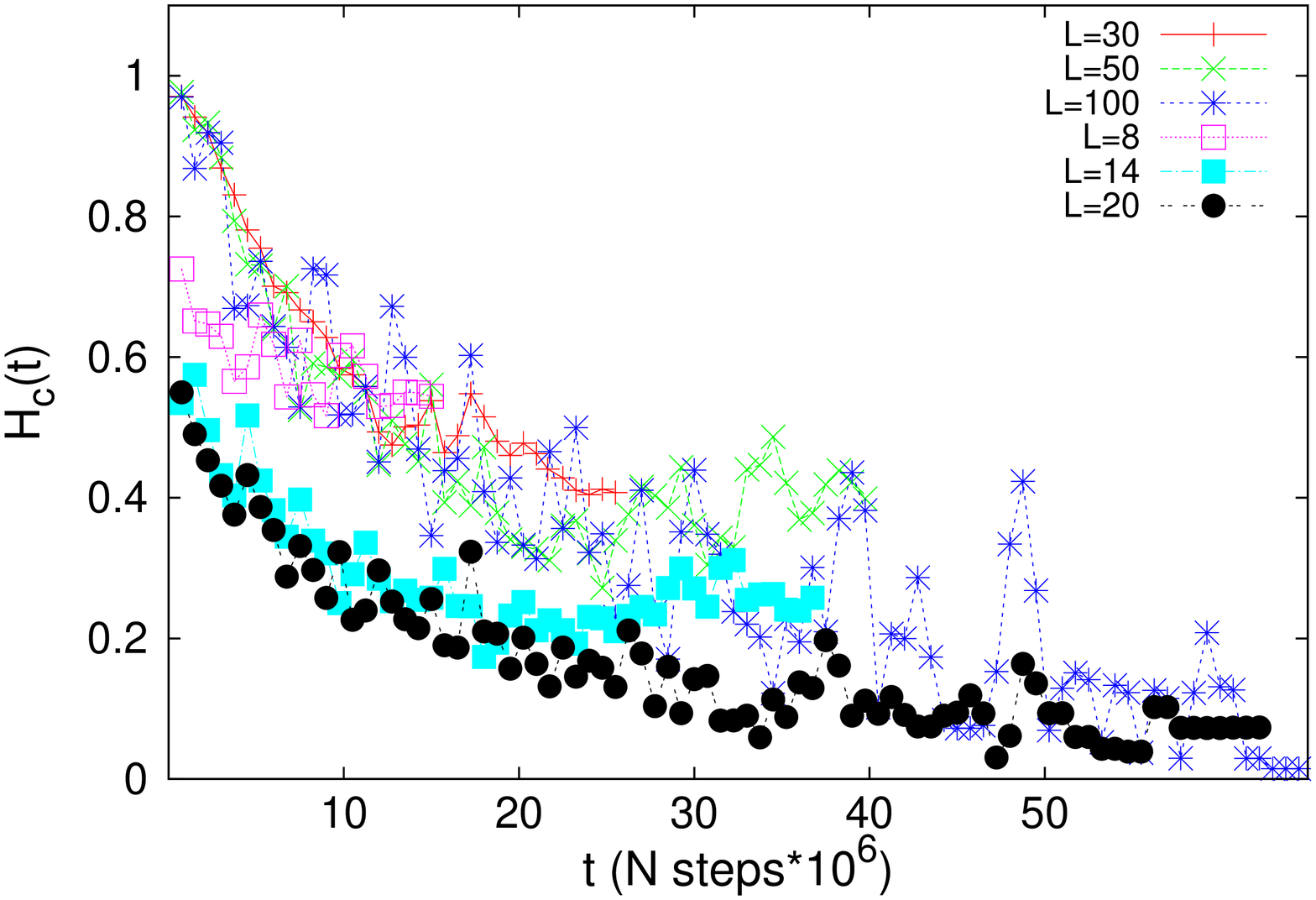} } & 
\rotatebox{-90}{\includegraphics[width=0.33\linewidth]{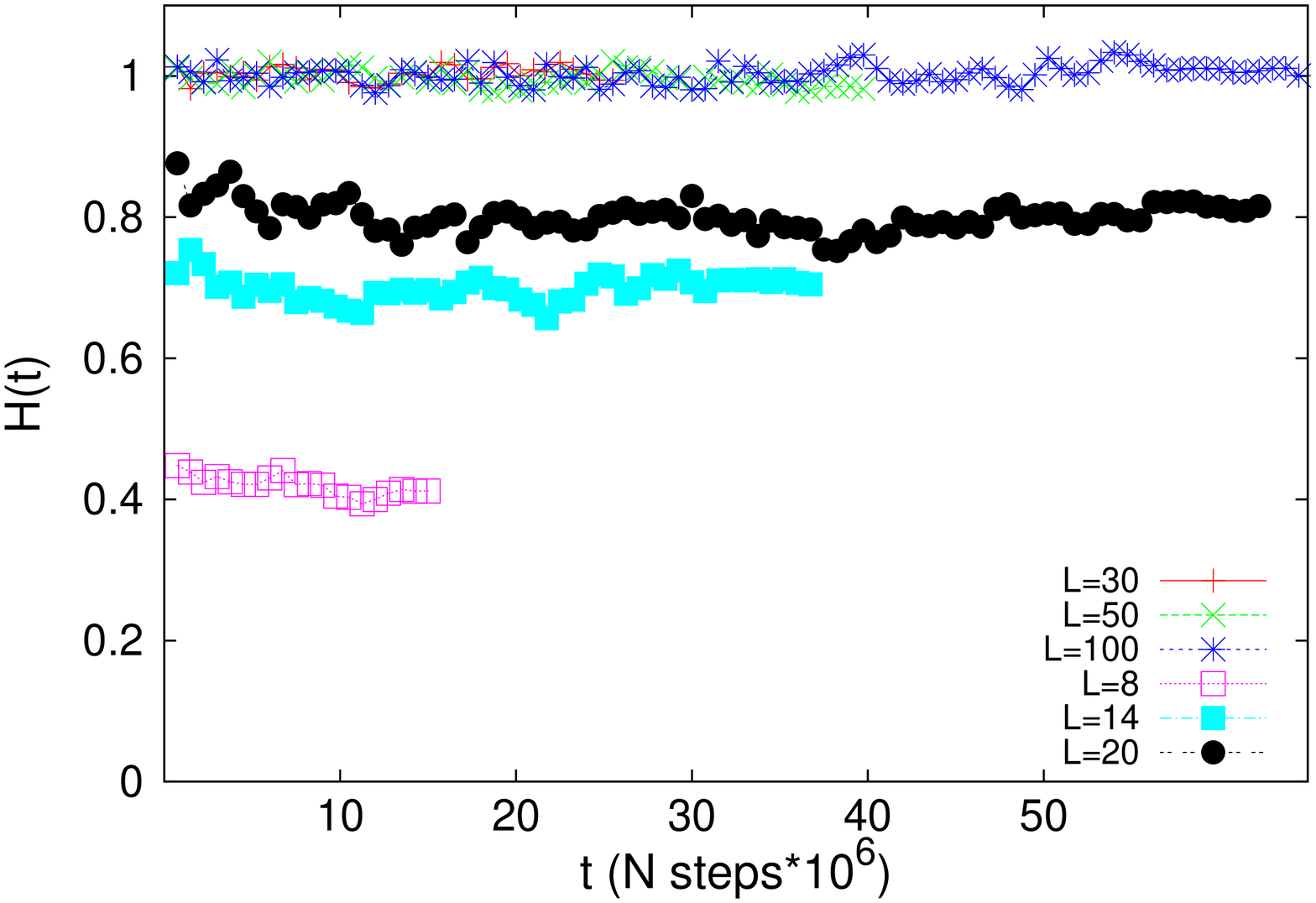}}
\end{array}
$ 
\caption{(Color online) Left: The Hamming distance between configurations 
corresponding to two consecutive minima of the temperature cycle is plotted vs. time.
Right:  The same but for the  distance between the initial configuration 
and that  corresponding to the current minimum.
All data are disorder averaged as explained in the main text.
}
\label{H_step} 
\end{figure}

The Hamming distance between configurations $\alpha$ and $\beta$ is
$H(\alpha,\beta) =1-\frac{1}{N_{spin}}\sum_i \sigma^\alpha_i \sigma^\beta_i$.
In the left panel of Fig.~\ref{H_step}, $\alpha$ denotes the configuration 
corresponding to the `current' BSF energy value, and $\beta$ is its immediate predecessor 
along a trajectory. In the right panel of the same figure, $\alpha$ is the initial configuration and
$\beta$ the current one. In other words, the left panel describes the configurational change 
occurring in a single thermal cycle, while the right panel describes the change accumulated from 
the beginning of the evolution of the system.
All data are disorder averaged as earlier explained.
In the 3d systems,  $H_c$, the Hamming distance between consecutive minima,
 has  a clear decreasing trend, interspersed by some oscillations.
Hence  the randomizing effect of crossing a record-sized barrier gradually 
tapers off, which is  a  desirable property in an optimization setting. The 2d case 
has much more pronounced oscillations, some of which correspond to 
system size configurational changes.

The right panel shows that, as expected, the Hamming distance 
of the current minimum  to the 
initial configuration nearly remains  constant in  time.
This  constant is near one in the 2d case,
implying  that low energy configurations successively identified are all 
nearly orthogonal to the initial configuration. In the 3d case the constant is
much smaller, indicating a higher degree of persistent correlation and a lingering memory of the past. 
In Fig.~\ref{Energia_step} 
the disorder averaged BSF energy per spin,  $e$, is plotted vs. time on a logarithmic horizontal scale
for 2d (left panel) and 3d systems.

\subsection{Comparing RDO and PT}
The RDO and PT algorithms are compared 
in terms of the disorder averaged BSF energy per spin,
respectively average energy per spin as a function of
temperature obtained using the two methods.
To simplify the notation, the same symbol $e$ is 
used for both quantities, both converging  for large times to the 
desired ground state energy, the target of the search.

We considered  two types of comparison: in the first, we use a `fast' PT,
where minimizing the execution time is a priority, but where the results
are in some  case of lesser quality than those obtained by RDO.
In the second, we use a `slow' version of PT algorithm, with parameters tuned
to obtain better, i.e. lower, energy values.
As mentioned,  slow PT requires  
the same computational  effort as  
 RDO, while fast PT  is 8 time faster than RDO.
All our results lie, as expected,  slightly above the thermodynamic 
limit of the model's ground state energy, see the previous model discussion  and the original 
references~\cite{Pal96,Roma09}.

\begin{figure}
 $
\begin{array}{cc}
\rotatebox{-90}{\includegraphics[width=0.33\linewidth]{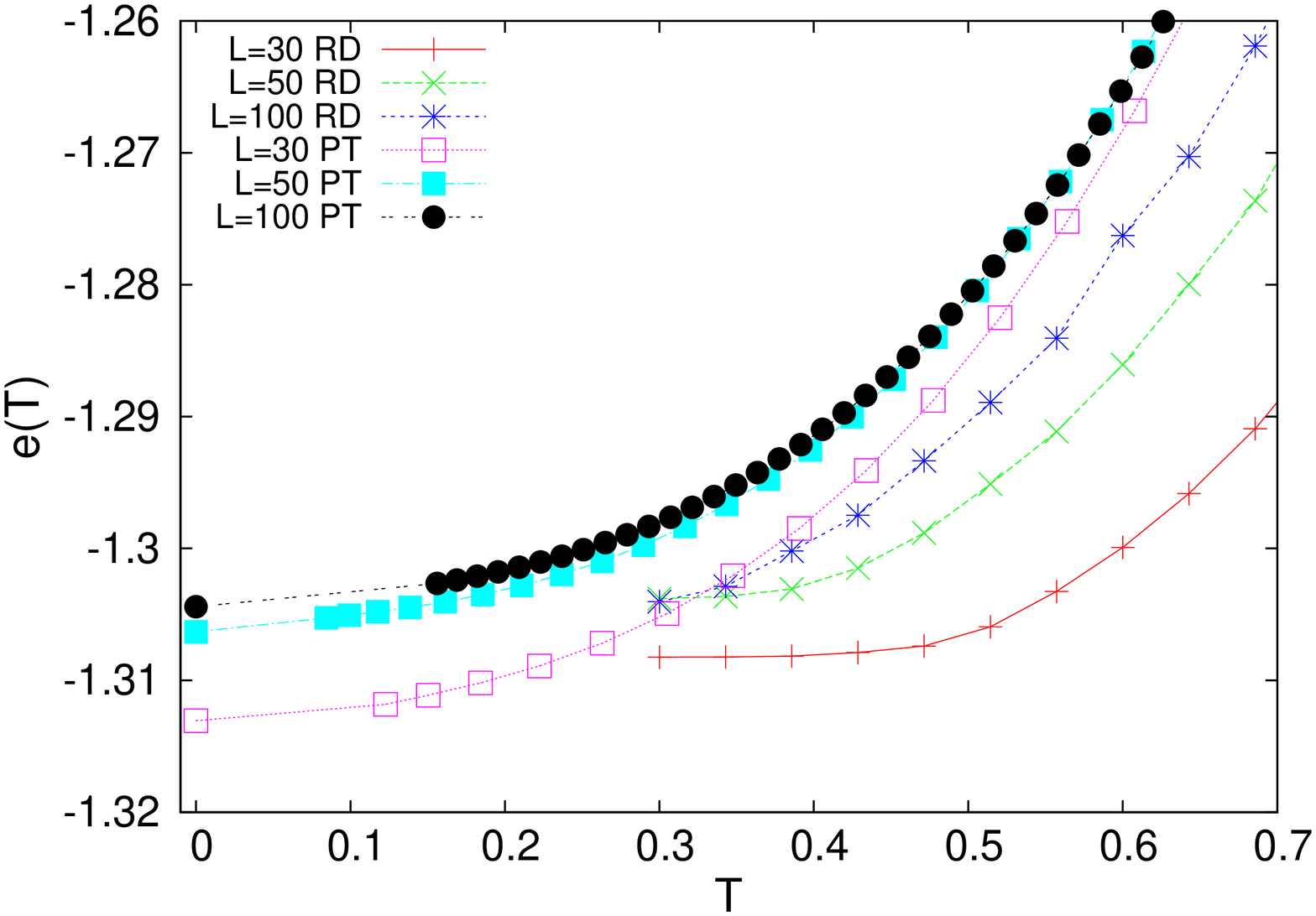}} & 
\rotatebox{-90}{\includegraphics[width=0.33\linewidth]{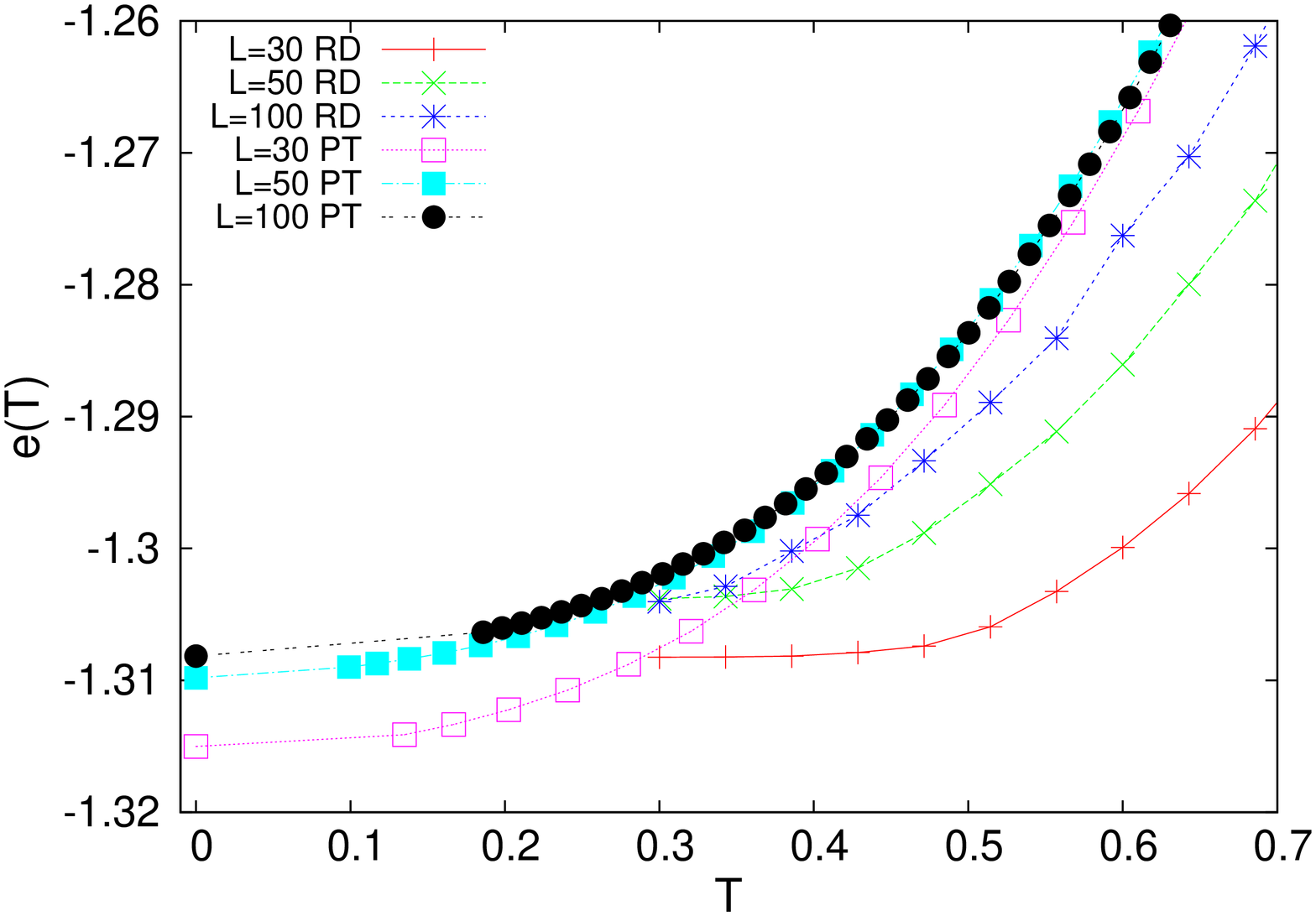}}
\end{array}
$ 
\caption{(Color online) Left: the BSF  energy, respectively average energy  per spin $e$ vs. temperature $T$ 
for all 2d systems considered. Data  obtained using  RDO and  fast  PT, respectively. 
Right: the same  RDO results compared with those  of  slow PT. 
}
\label{Energia2d_T} 
\end{figure}

In the left panel of Fig.~\ref{Energia2d_T},  the average BSF energy per spin $e$ of the 
2D system  is plotted as a function of the temperature $T$. As explained in the figure text,
the  RDO  and fast PT results are plotted for different system sizes. Interestingly, the average energy 
calculated by the RDO is consistently lower than its PT counterpart. 

Also note that BSF energy   obtained in the RDO at $T=0.3.$  is comparable or,
 in the case of the larger systems, even lower than the average energy obtained in PT at a lower 
temperature.
 On the right panel we compare  RDO with  our  slow PT.
Here    PT finds  lower energies than RDO, but  
the difference is hardly  significant.

\begin{figure}
 $
\begin{array}{cc}
\rotatebox{-90}{\includegraphics[width=0.33\linewidth]{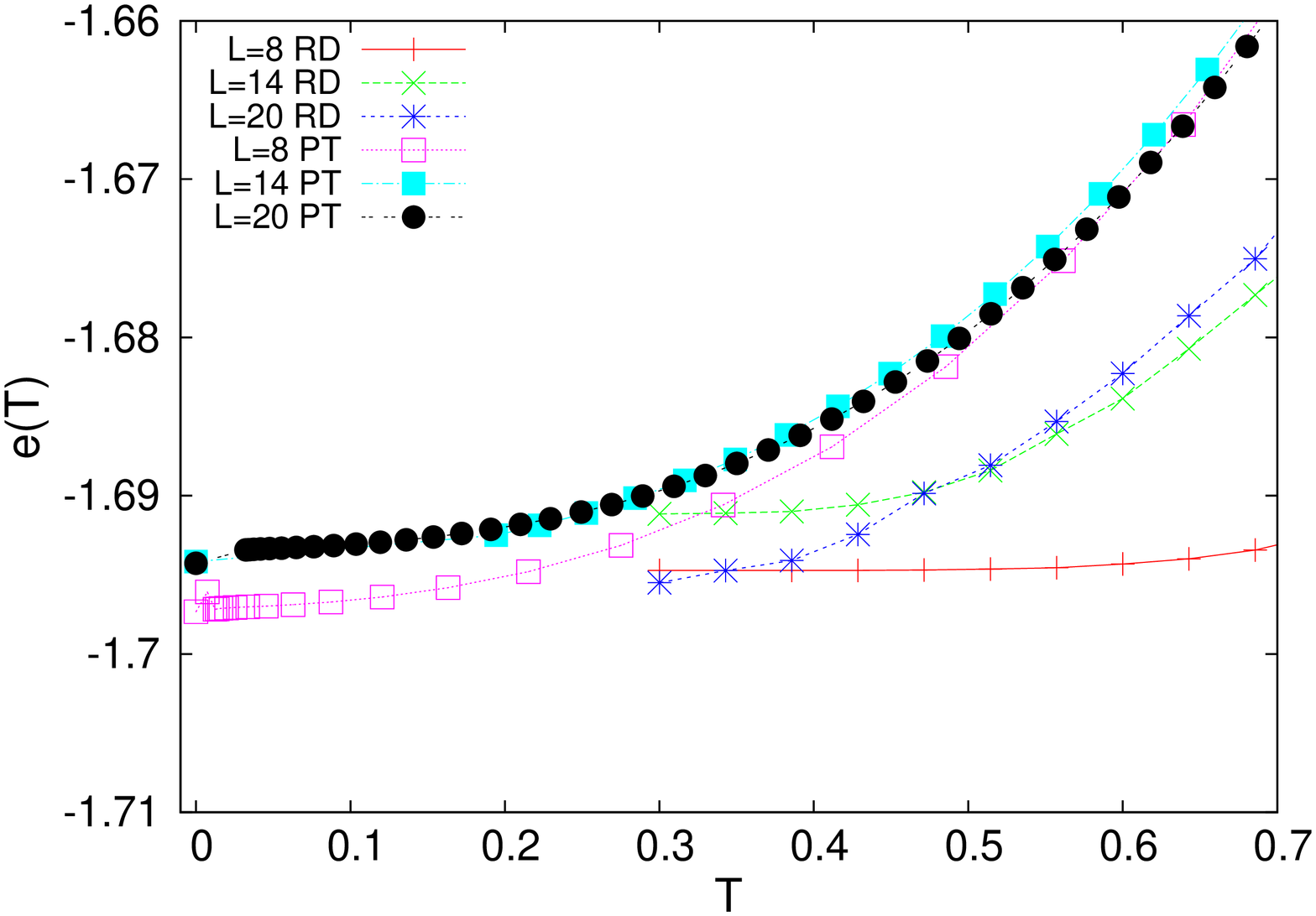}} & 
\rotatebox{-90}{\includegraphics[width=0.33\linewidth]{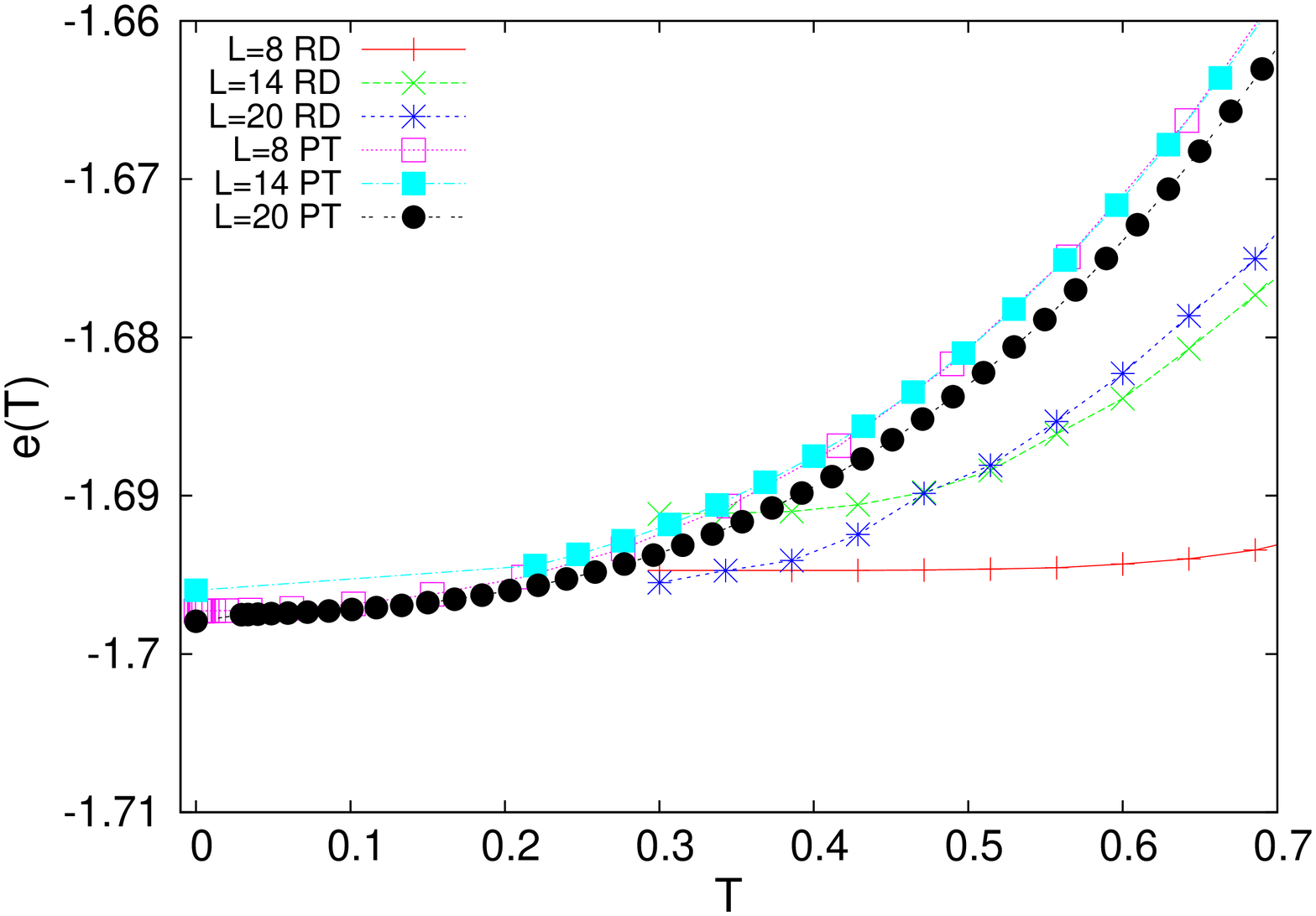}}\\
\end{array}
$ 
\caption{ (Color online)
Left: the BSF  energy, respectively average energy  per spin $e$ 
for all 3d systems considered. Data  obtained using  RDO  and fast  PT, respectively. 
Right: the same  RDO results compared with those  of  slow PT. 
}
\label{Energia3d_T} 
\end{figure} 

Figure~\ref{Energia3d_T} is the 3-d analog  Fig.~\ref{Energia2d_T} 
 and similar observations apply to  the trends and  minima.
Furthermore, the BSF energy given by   RDO are almost coincident with the ground states at $T=0$ given by  
slow PT.
\begin{table}[ht]
  \centering
  \begin{tabular}{c c c c}
    \hline \hline
   & RDO & Fast PT & Slow PT \\
    \hline
L=30 & -1.30824$\pm$0.00167 & -1.31306$\pm$0.00166& -1.31501$\pm$0.00166\\
    \hline
L=50 & -1.30380$\pm$0.00202 & -1.30632$\pm$0.00200& -1.30981$\pm$0.00202\\
    \hline
L=100 &-1.30403$\pm$0.00214 & -1.30441$\pm$0.00212& -1.30815$\pm$0.00217\\
    \hline
L=8   &-1.69472 $\pm$0.00236 &-1.69728$\pm$0.00236& -1.69733$\pm$0.00276\\
    \hline
L=14  &-1.69114$\pm$0.00194  &-1.69416$\pm$0.00191& -1.69597$\pm$0.00188\\
\hline 
L=20  &-1.69549$\pm$0.00194  &-1.69426$\pm$0.00214& -1.69793$\pm$0.00215\\
\hline 
\hline
  \end{tabular}
  \caption{Ground-state energy estimates (energy per spin)
obtained using  RDO and the fast and slow PT algorithm.
The first three rows pertain to 2d systems, and the last three to 3d systems.
The errors are given as $\pm \sigma $, where $\sigma$ is the 
estimated standard error on 
   the computed averages.}
  \label{ground state}
\end{table}
Table~\ref{ground state} summarizes the estimated ground state enenrgy values obtained for 
all the systems considered using RDO and our fast and slow PT algorithm.
The values quated are ensable averages and their standard
($1 \sigma$) errors.
The two methods very nearly produce results with overlapping
error bars. The fast  PT 
algorithm results for the largest systems are marginally inferior to the corresponding
RDO results, while the slow PT results are marginally superior.

\section{Discussion}
Similarly to Simulated Annealing (SA),  Record Dynamics Optimization is a `thermal' 
optimization heuristics based on local search and on the
 Metropolis acceptance rule.
  Unlike SA, it  features  an alternation of  
heating and cooling phases, each  delimited  by  the achievement of a  record high `barrier energy',
and   a lower Best So Far (BSF) energy, respectively.  The current BSF energy 
provides an estimate of the solution of the optimization
problem at hand.
The physical idea
behind RDO  is that  the configuration space of hard optimization  problems explored by
local searches  can be coarse-grained into
a hierarchy of nested sets. Starting from a poor solution, 
configurations of decreasing cost
 can only be  accessed by scaling  increasingly large barriers. Hence, quickly generating record high barriers provides a effective way to
achieve better solutions.  For a given application, the validity of a hierachical  description
  is buttressed 
whenever RDO   works efficiently. In this way, which makes RDO into 
a landscape optimization tool.

RDO has a modicum of adjustable parameters. Most important are  the cooling/heating rate,
and the number of BSF energy values found  when (briefly) searching at constant temperature
following a  cooling phase. The programming effort in  RDO is similar to standard SA and
considerably smaller  than in  PT.
On our  test problem,  RDO  seems to deliver  marginally higher  energy values  than     
the slow PT algorithm on the largest systems considered.
However, PT is a highly optimized algorithm with a long history of 
successes, while RDO is a new algorithm, which, we surmise, still has 
 considerable  potential for further improvements.
 
\section{Acknowledgments} DB is grateful to Federico Ricci-Tersenghi for introducing him to 
computational physics in general and to Parallel Tempering in particular.

\bibliographystyle{unsrt}
\bibliography{references,SD-meld}

\begin{thebibliography}{10}

\bibitem{Kirkpatrick83}
S.~Kirkpatrick, C.D.~Gelatt Jr., and M.~P. Vecchi.
\newblock Optimization by simulated annealing.
\newblock {\em Science}, 220:671--680, 1983.

\bibitem{Cerny85}
V.~{\u{C}}ern{\'{y}}.
\newblock Thermodynamical approach to the traveling salesman problem: An
  efficient simulation algorithm.
\newblock {\em Journal of Optimization Theory and its Applications}, 45:41--55,
  1985.

\bibitem{Huang86}
M.~D. Huang, Fabio Romeo, and Alberto Sangiovanni-{V}incentelli.
\newblock An efficient general cooling schedule for simulated annealing.
\newblock In {\em International Conference on Computer Aided Design}, page 381,
  1986.

\bibitem{Goldberg89}
David~E. Goldberg.
\newblock {\em Genetic algorithms in search, optimization and machine
  learning}.
\newblock Addison-Wesley, Reading, Massachusset, 1989.

\bibitem{Dueck90}
G.~Dueck and T.~Scheuer.
\newblock Threshold accepting: A general purpose optimization algorithm
  appearing superior to simulated annealing.
\newblock {\em Journal of Computational Physics}, 90:161--175, 1990.

\bibitem{Nourani99}
Y.~Nourani and B.~Andresen.
\newblock Exploration of np-hard enumeration probØlems by simulated annealing
  - the spectrum values of permanents.
\newblock {\em Theoretical Computer Science}, 215:51, 1999.

\bibitem{boettcher01}
S.~Boettcher and A.~Percus.
\newblock Optimization with extremal dynamics.
\newblock {\em Phys. Rev. Lett.}, 86:5211--5214, 2001.

\bibitem{vanLaarhoven87}
P.~J.~M. van Laarhoven and E.~H.~L. Aarts.
\newblock {\em Simulated annealing}.
\newblock D.~Reidel Publishing Company, Dordrecht, 1987.

\bibitem{Salamon02}
Peter Salamon, Paolo Sibani, and Richard Frost.
\newblock {\em Facts, conjectures and improvements for simulated annealing}.
\newblock SIAM, Philadelphia, 2002.

\bibitem{Sibani03}
Paolo Sibani and Jesper Dall.
\newblock {Log-Poisson statistics and pure aging in glassy systems.}
\newblock {\em Europhys. Lett.}, 64:8--14, 2003.

\bibitem{Anderson04}
{Paul Anderson, Henrik Jeldtoft Jensen, L.P. Oliveira and Paolo Sibani}.
\newblock Evolution in complex systems.
\newblock {\em Complexity}, 10:49--56, 2004.

\bibitem{Sibani06a}
{Paolo Sibani, G.F. Rodriguez and G.G. Kenning}.
\newblock Intermittent quakes and record dynamics in the thermoremanent
  magnetization of a spin-glass.
\newblock {\em Phys. Rev. B}, 74:224407, 2006.

\bibitem{Sibani08}
Paolo Sibani and Simon Christiansen.
\newblock {Thermal shifts and intermittent linear response of aging systems}.
\newblock {\em Phys. Rev. E}, {77}({4, Part 1}), {APR} {2008}.

\bibitem{Edwards75}
S.~F. Edwards and P.~W. Anderson.
\newblock Theory of spin glasses.
\newblock {\em J. Phys. F}, 5:965--974, 1975.

\bibitem{Dall01}
Jesper Dall and Paolo Sibani.
\newblock {Faster} {M}onte {C}arlo simulations at low temperatures. {T}he
  waiting time method.
\newblock {\em Comp. Phys. Comm.}, 141:260--267, 2001.

\bibitem{Hoffmann88}
Karl~Heinz Hoffmann and Paolo Sibani.
\newblock Diffusion in hierarchies.
\newblock {\em Phys. Rev. A}, 38:4261--4270, 1988.

\bibitem{Mobius97}
A.~M{\"o}bius, A.~Neklioudov, A.~D{\'i}az-S{\'a}nchez, K.~H. Hoffmann,
  A.~Fachat, and M.~Schreiber.
\newblock Optimization by {T}hermal {C}ycling.
\newblock {\em Phys. Rev. Lett.}, 79:4297--4301, 1997.

\bibitem{Marinari}
I.~Kondor~(Eds.) E.~Marinari, in: J.~Kertesz.
\newblock {\em {Optimized Monte Carlo Methods, Lectures given at the 1996
  Budapest Summer School on Monte Carlo Methods}}.
\newblock Springer-Verlag, 1996.

\bibitem{Hukushima}
K.~Hukushima and K.~Nemoto.
\newblock {Exchange Monte Carlo Method and Application to Spin Glass
  Simulations.}
\newblock {\em J. Phys. Soc. Jpn.}, 65, 1996.

\bibitem{Moreno}
H.G.~Katzgraber J.J.~Moreno and A.K. Hartmann.
\newblock {Finding Low-Temperature States with Parallel Tempering, Simulated
  Annealing and Simple Monte Carlo.}
\newblock {\em International Journal of Modern Physics C}, 14:285, 2003.

\bibitem{Marinari2}
E.~Marinari, G.~Parisi, F.~Ricci-Tersenghi, J.~Ruiz-Lorenzo, and F.~Zuliani.
\newblock {Replica Symmetry Breaking in Short Range Spin Glasses: A Review of
  the Theoretical Foundations and of the Numerical Evidence. }.
\newblock {\em J. Phys.}, 98:973, 2000.

\bibitem{Pal96}
Karoly~F. Pal.
\newblock The ground state of the cubic spin-glass with short-range
  interactions of gaussian distribution.
\newblock {\em Physica A}, 233:60--66, 1996.

\bibitem{Roma09}
F.~Rom{\'{a}}, S.~Risau-Gusman, A.J. Ramirez-Pastor, F.~Nieto, and E.E. Vogel.
\newblock {The ground state energy of the Edwards-Anderson spin glass model
  with a parallel tempering Monte Carlo algorithm}.
\newblock {\em Physica A}, 388:2821--2838, 2009.

\bibitem{Dall03}
Jesper Dall and Paolo Sibani.
\newblock Exploring valleys of aging systems: the spin glass case.
\newblock {\em Eur. Phys. J. B}, 36:233--243, 2003.

\end{thebibliography}
\end{document}